\documentclass[aps,amsfonts,superscriptaddress, nofootinbib, floatfix, 10pt,twocolumn,prd,showpacs,showkeys]{revtex4-1}
\usepackage{graphicx}
\usepackage{amsmath}
\usepackage[latin1]{inputenc}
\usepackage{amsbsy}
\usepackage{color}
\usepackage{epsfig}
\usepackage{graphicx}
\usepackage{amsmath}
\usepackage{amssymb}
\usepackage{bbm}
\usepackage{amsbsy}
\usepackage{slashed}
\DeclareMathOperator{\tr}{tr}

\usepackage{xcolor}

\newcommand{\beq}{\begin{equation}}
\newcommand{\eeq}{\end{equation}}

\definecolor{mbcol}{rgb}{1,0,1}

\usepackage[normalem]{ulem}  

\begin{document}
\title{The dual quark condensate in local and nonlocal NJL models:
an order parameter for deconfinement?}

\author{Federico Marquez}
\affiliation{Instituto de F\'isica, Pontificia Universidad Cat\'olica de Chile, Casilla 306, Santiago 22, Chile.}
\author{Aftab Ahmad}
\affiliation{Instituto de F\'{\i}sica y Matem\'aticas, Universidad Michoacana de San Nicol\'as de Hidalgo. Edificio C-3, Ciudad Universitaria, Morelia 58040, Michoac\'an, M\'exico.}
\affiliation{Department of Physics, Gomal University.  D.I. Khan 29220, K.P.K., Pakistan.}
\author{Michael Buballa}
\affiliation{Theoriezentrum, Institut f\"ur Kernphysik, Technische Universit\"at Darmstadt, 
Schlossgartenstr.~2, 64289 Darmstadt, Germany}
\author{Alfredo Raya}
\affiliation{Instituto de F\'isica, Pontificia Universidad Cat\'olica de Chile, Casilla 306, Santiago 22, Chile.}
\affiliation{Instituto de F\'{\i}sica y Matem\'aticas, Universidad Michoacana de San Nicol\'as de Hidalgo. Edificio C-3, Ciudad Universitaria, Morelia 58040, Michoac\'an, M\'exico.}

\begin{abstract}

We study the behavior of the dual quark condensate $\Sigma_1$ 
in the Nambu--Jona-Lasinio (NJL) model and its nonlocal variant.
In quantum chromodynmics $\Sigma_1$ can be related to the breaking of the center symmetry and is therefore an (approximate) 
order parameter of confinement. 
The deconfinement transition is then signaled by a strong rise of $\Sigma_1$ as a function of temperature. However, a similar behavior is also seen in the NJL model, which is known to have no confinement.
Indeed, it was shown that in this model the rise of $\Sigma_1$ is triggered by the {\it chiral} phase transition.
In order to shed more light on this issue,
we calculate $\Sigma_1$ for several variants of the NJL model, some of which have been suggested to be confining. 
Switching between ``confining'' and ``non-confining'' models and parametrizations we find no qualitative difference 
in the behavior of $\Sigma_1$, namely, it always rises in the region of the chiral phase transition.
We conclude that without having established a relation to the center symmetry in a given model, 
$\Sigma_1$ should not blindly be regarded as an order parameter of confinement.

\end{abstract}
\maketitle

\section{Introduction}

Spontaneous chiral symmetry breaking and confinement are among the most important features of 
nonperturbative low-energy quantum chromodynamics (QCD). 
Both properties have been studied from first principles within lattice QCD, 
which nowadays provides an excellent description of the meson and baryon spectrum in vacuum~\cite{Durr:2008zz}.
Turning to nonzero temperature $T$, lattice QCD predicts a rapid but smooth crossover from the 
confined phase with broken chiral symmetry at low $T$ to a deconfined phase with (approximately)
restored chiral symmetry at high $T$~\cite{Borsanyi:2010bp,Bazavov:2011nk}. 
Moreover, it is found that both transitions take place in the same temperature range.
The corresponding order parameters are the chiral quark condensate  $\langle\bar{\psi}\psi\rangle$
and the Polyakov-loop (PL) expectation value~\cite{Polyakov:1978vu,Susskind:1979up}, respectively, 
which strictly speaking belong to opposite limits of the theory.   
While $\langle\bar{\psi}\psi\rangle$ is a strict order parameter for chiral symmetry breaking in the limit of massless quarks,
the  PL expectation value is a strict order parameter for confinement in the limit of infinitely heavy quarks,  
where it can be related to $\mathbb{Z}(3)$ center-symmetry breaking and thereby to the free energy of 
a static quark-antiquark pair~\cite{McLerran:1980pk,Karsch:1985cb}.

The identification of an appropriate order parameter for the confinement-deconfinement phase transition 
in QCD with finite (i.e., non-infinite) quark masses stands on less solid grounds.
In this context the authors of Ref.~\cite{EBF:08} have proposed the dual quark condensate $\Sigma_1$ as an alternative.
Starting point is the generalized quark condensate $\langle\bar{\psi}\psi\rangle_\phi$, which is the analogue
of the usual quark condensate, but evaluated 
for quark fields with twisted boundary conditions 
\begin{equation}
\psi(\vec x,\beta)={\rm e}^{-i\phi}\psi(\vec x,0)
\label{twboundary}
\end{equation}
in the imaginary time direction.
Here  $\beta = 1/T$ is the inverse temperature
and
 $\phi \in [0,2\pi)$ is an arbitrary angle.
 Hence, physical fermions, which obey antiperiodic boundary conditions, correspond to $\phi = \pi$.

The dual quark condensate $\Sigma_n$ is then defined as the Fourier transform with respect to $\phi$,
\begin{equation}
\Sigma_n=-\int^{2\pi}_{0}\frac{d\phi}{2\pi} e^{-i n\phi}\langle\bar{\psi}\psi\rangle_{\phi},
\label{Sigman}
\end{equation}
where $n$ is an integer.
In lattice representation this can be written as a sum of Wilson loops winding $n$ times around  
the temporal boundary~\cite{EBF:08}. 
In particular, since the PL is the shortest loop with winding number 1, the case $n=1$ 
may be viewed as a collection of generalized Polyakov loops with spatial fluctuations.
Therefore $\Sigma_1$  has been termed ``dressed Polyakov loop'' \cite{EBF:08}.

$\Sigma_1$ and the ordinary (``thin'') PL transform in the same manner under center
transformations, which motivates the consideration of $\Sigma_1$ as an order parameter for the  
deconfinement phase transition.
Moreover, since the spatial fluctuations are suppressed for infinite quark masses, $\Sigma_1$ reduces to the 
thin PL in this limit. On the other hand, as seen from its definition, it is also related to the quark condensate, 
albeit with unphysical boundary conditions. 
This hints for a possible  connection between chiral and deconfinement phase transition, 
explaining why both transitions occur in the same temperature region~\cite{EBF:08,Braun:2009gm}. 

Another important feature of $\Sigma_1$ is that it is not restricted to lattice formulations of QCD,
but  it can also be calculated within continuum approaches, like the functional renormalization group method~\cite{Braun:2009gm}
or Schwinger-Dyson equations~\cite{CSF:09,CSFM:09,CSFA:10,Fischer:2011mz,Fischer:2014ata},
where the calculation of the thin Polyakov loop is not possible in a straightforward manner. 
These investigations have confirmed that chiral restoration and deconfinement phase transitions take place in the same
regime, even when the analysis is extended to a nonvanishing chemical potential~\cite{Fischer:2011mz,Fischer:2014ata}, 
a region which is not accessible in lattice QCD because of the sign problem. 

In addition to these studies directly rooted in QCD, $\Sigma_1$  has also been investigated within effective models of 
strong interactions. 
This was done first in the Polyakov-loop extended NJL (PNJL) model~\cite{KKH:09} and a bit later in the 
standard NJL model~\cite{TKM:10}. 
It was found in both cases that the qualitative behavior is similar to the QCD results,
i.e., the rise of $\Sigma_1$, which in QCD signals the onset of deconfinement, is found approximately in the same temperature 
region where chiral restoration takes place. 
This is particularly remarkable for the NJL model, which has been invented as a model for chiral symmetry 
breaking~\cite{Nambu:1961tp,Nambu:1961fr}
but is known to feature no confinement. 
Indeed, since there are no gauge fields and, hence, no center symmetry in this model, a connection between $\Sigma_1$ and confinement cannot be made.\footnote{
A similar conclusion was drawn in Ref.~\cite{Feng:2013bsa}, where $\Sigma_1$ has been explored in three-dimensional quantum electrodynamics, a confining theory which, however,  does not possess a center symmetry. 
Therefore, although the behavior of $\Sigma_1$ is qualitatively the same as in QCD, 
it cannot directly be linked to the confinement transition.
}
Instead, a Ginzburg-Landau type analysis revealed 
that in the NJL model the behavior of $\Sigma_1$ is triggered by the chiral phase transition~\cite{Benic01}.

The situation is somewhat less clear for the PNJL model, where confinement effects are included by coupling 
a gluon background to the quarks~\cite{Meisinger:1995ih,Fukushima:2003fw}. 
However, in this way confinement is realized only statistically, i.e., quark effects to thermodynamic quantities are
suppressed at low temperature, but  the quarks remain as physical states in the spectrum.
As a consequence mesons can decay into quark-antiquark pairs~\cite{Hansen:2006ee},
which should not be possible in a confining theory.

Already before the PNJL model was invented,
there have been various attempts to model confinement by modifying the analytic properties
of the quark propagator, e.g.~\cite{Krein:1990sf,Burden:1991gd,Stainsby:1992hy,Gribov:1999ui}. 
One possibility is a quark propagator without real singularities in the time-like momentum region. 
Formally, this is related to the violation of reflection positivity~\cite{Roberts:2000aa, Alkofer:2000wg},
meaning that quarks do not exist as physical states in the particle spectrum and are thus confined.

As mentioned above, the NJL model in its original formulation does not support confinement; 
the structure of the quark propagator in this model is consistent with that of a free particle. 
However, 	employing the proper-time regularization, the unphysical quark production threshold can be avoided by 
introducing  an infrared cutoff, associated with the confining scale~\cite{Ebert:1996vx,Hellstern:1997nv}.
The resulting propagator does neither develop real nor complex poles.
This is another statement of confinement in the sense that the excitation described by a pole-less propagator 
can never reach its mass shell.

Alternative attempts to simulate confinement utilize nonlocal extensions of the NJL model (nNJL model)~\cite{Buballa:1992sz,Bowler:1994ir,Plant:1997jr,General:2000zx}.
Thereby the interaction is designed in such a way that the quark propagator has no real but complex conjugate poles.\footnote{
An exception is the propagator in  \cite{Buballa:1992sz}, which has no pole in the complex momentum plane
but complex conjugate cuts.}
These complex singularities have been interpreted as confined quasiparticles \cite{Yo01, Yo02}. Through the incorporation of temperature into the model, the complex singularities may turn real and a deconfinement phase transition becomes explicit.

It should be noted that the absence of real poles in the quark propagator
is a sufficient but not a necessary criterion for confinement.
In fact, in contrast to older studies~\cite{Roberts:2000aa}, the Schwinger-Dyson analysis of \cite{Alkofer:2003jj} seems 
to favor the existence of a real quark pole when a truncation scheme beyond the rainbow-ladder approximation is used. 
It was also pointed out that the existence of complex conjugate poles in the nNJL model leads to thermodynamic
instabilities~\cite{Benic:2012ec} and only inhibits the decay of mesons into quark-antiquark pairs  
if additional prescriptions about the integration contour are made~\cite{Plant:1997jr}.

Nevertheless these models provide a nice and relatively simple test bed to investigate
whether the pole structure of the quark propagator  (``confining'' or ``non-confining'' in the above sense) 
has a qualitative effect on the behavior of $\Sigma_1$.
In this article, we therefore explore the behavior of $\Sigma_1$ in the NJL and nNJL models such that, with the appropriate 
choice of parameters, confinement is built-in or not in the models.
Thereby we address the validity of $\Sigma_1$ as an order parameter for the confinement-deconfinement transition in these models.

The remainder of this article is organized as follows: 
In Sec.~\ref{sec:models} we introduce the NJL and nNJL models 
and explain how the chiral condensate and $\Sigma_1$ are obtained within these models.
The corresponding results are presented in Sec.~\ref{sec:results}.
Finally, in Sec.~\ref{sec:conclu}, we draw our conclusions.

\section{The models}~\label{sec:models}

\subsection{Local NJL model with infrared cutoff}
\label{sec:PTRNJL}

The Nambu--Jona-Lasinio (NJL) model~\cite{Nambu:1961tp,Nambu:1961fr} 
is a model of self-interacting fermions and was introduced 
in the early 1960's to describe the mass of nucleons through spontaneous chiral symmetry breaking. 
After the  advent of QCD it was reinterpreted as an effective model for quarks, which acquire a constituent 
mass by the same mechanism (see 
Refs.~\cite{Vogl:1991qt,Klevansky01,Hatsuda:1994pi,Buballa01} for reviews).

Throughout this article we work in Euclidean space, following the conventions of Ref.~\cite{Scoccola01}. 
The NJL model is then given by the Lagrangian
\begin{equation}
\mathcal{L}_\mathrm{NJL}=
\bar{\psi}(-i\slashed{\partial}+m)\psi
-\frac{G}{2}\Bigg((\bar{\psi}\psi)^2+(\bar{\psi}i\gamma^5\boldsymbol{\tau}\psi)^2\Bigg),
\label{LNJL}
\end{equation}
where $\psi$ denotes a quark field with $N_f=2$ flavor and $N_c=3$ color degrees of freedom,
and with bare mass $m$.
The operator $\slashed{\partial}$ is defined as
\beq
       \slashed{\partial} = \gamma_4 \frac{\partial}{\partial x_4} + \boldsymbol{\gamma}\cdot\boldsymbol{\nabla}\;,
\eeq
with $\gamma_4 = i\gamma^0$ and the imaginary time variable $x_4 = ix^0 \equiv it$. 
The quarks interact by local four-point vertices, proportional to the coupling constant $G$.
The interaction is invariant under chiral $SU(2)_L\times SU(2)_R$ transformations and
consists of a scalar-isoscalar and a pseudoscalar-isovector part, where $\boldsymbol{\tau}$ 
denotes the triplet of the Pauli matrices in isospin space.

In vacuum chiral symmetry, which is explicitly broken by the bare mass $m$, is also broken spontaneously 
through the self-interactions. 
In one-loop approximation (being equivalent to the mean-field or Hartree approximation), this gives rise
to the dressed Euclidean quark propagator
\begin{equation}
S(q)=\frac{-\slashed{q}+M}{q^2+M^2},
\label{Sdressed}
\end{equation}
where $M$ denotes the constituent quark mass,
given by the gap equation
\begin{equation}
M=m+G\int\frac{d^4q}{(2\pi)^4}\tr (S(q)).
\label{gapNJL1}
\end{equation}
Here the trace is to be taken over the internal quark degrees of freedom, i.e., Dirac, color and flavor components.
One finds 
\begin{equation}
M = m + 8N_c G\int\frac{d^4q}{(2\pi)^4}\frac{M}{q^2+M^2}.
\label{gapNJL2}
\end{equation}
Since the constituent quark mass also enters the right-hand side,  the equation 
must be solved self-consistently, underlining its non-perturbative nature.

The quark condensate  is generally given by the expression
\begin{equation}
\langle\bar{\psi}\psi\rangle = -\int\frac{d^4q}{(2\pi)^4}\tr (S(q)) .
\label{qqgen}
\end{equation}
If we compare this with Eq.~(\ref{gapNJL1}) we find the simple relation
\beq
        \langle\bar{\psi}\psi\rangle = -\frac{M-m}{G}.
\label{pbpM}
\eeq

Applying Matsubara formalism, the above expressions can straightforwardly be generalized to nonzero temperature 
by the substitutions
\begin{eqnarray}
\int\frac{dq_4}{2\pi}&\rightarrow&T\sum_n \; ,
\label{FTsub1}
\\
q_4&\rightarrow&\omega_n
\;,
\label{FTsub2}
\end{eqnarray}
where $\omega_n = (2n+1)\pi T$ are  fermionic Matsubara frequencies. 
Furthermore,
recalling that the latter are a consequence of the antiperiodic boundary conditions
fermions must obey in the imaginary time direction, 
the twisted boundary conditions defined in Eq.~(\ref{twboundary}) are easily implemented by the shift 
\begin{equation}
\omega_n\rightarrow\omega_{n}^{\phi}=\left(2n+\frac{\phi}{\pi}\right)\pi T.
\label{omeganphi}
\end{equation}

So far we have ignored the fact that the integrals in Eqs.~(\ref{gapNJL1}) -- (\ref{qqgen}) 
and their extensions to nonzero temperature and twisted boundary conditions  
diverge in the ultraviolet and therefore have to be regularized. Since the NJL model is not renormalizable, 
the divergences cannot be absorbed in a redefinition of the parameters in the Lagrangian,
so that the regularization scheme and the corresponding cut-off parameters must be viewed as 
a part of the model.

As already mentioned, a second shortcoming of the NJL model  in its original form
is the fact that the model is not confining. 
Formally this can be seen most easily from the dressed propagator, Eq.~(\ref{Sdressed}), 
which takes the form of the propagator of a non-interacting particle.
In particular it has a pole in the time-like region at $q^2 = -M^2$, 
corresponding to the mass shell of a real particle.

It is possible, however, to circumvent this problem by choosing a regularization scheme 
which avoids the poles of the propagator and, hence, the appearance of quark-production
thresholds~\cite{Ebert:1996vx,Hellstern:1997nv}.
The basic idea is to restrict the distance the confined quark can propagate by introducing
an infrared cut-off, in addition to the one needed to regularize the UV divergences. 
The proper-time (PT) regularization turned out to be best suited for this task.
To this end we use the identity
 \begin{equation}
\frac{1}{q^2+M^2}=\int^{\infty }_{0} d\tau \, e^{-\tau(q^2+M^2)}
\end{equation} 
to write the propagator as an integral. 
If we now restrict the domain of the integration by introducing an IR cut-off $\tau_\mathit{ir}^2$ 
and  a UV one $\tau_\mathit{uv}^2$, we find
\beq
\int^{\tau_\mathit{ir}^2}_{\tau_\mathit{uv}^2} \!\!\! d\tau \, e^{-\tau(q^2+M^2)}
=\frac{ e^{-\tau_\mathit{uv}^2(q^2+M^2)}-e^{-\tau_\mathit{ir}^2(q^2+M^2)}}{q^2+M^2},
\label{PTRc}
\eeq
where the original pole at $q^2 = -M^2$  is canceled by the numerator.
Hence, we have removed the singularities from the propagator.
As discussed before, this can be interpreted as confinement. 

Applying this scheme to Eq.~(\ref{gapNJL2}), the gap equation in vacuum becomes~\cite{Ebert:1996vx}
\begin{equation}
M = m 
+\frac{N_c}{2\pi^2}\,GM^3 \,\Big(\Gamma(-1,M^2\tau^2_\mathit{uv})
-\Gamma(-1,M^2\tau^2_\mathit{ir}) \Big),
\label{PTRdnew}
\end{equation}
where $\Gamma(\alpha,x)$ is the incomplete gamma function.
At nonzero temperature one gets 
\begin{align}
M = m - \frac{2N_c}{\pi^{3/2}} GMT \sum_n
\Bigg[\;  \frac{e^{-M_n^{2}\tau_\mathit{uv}^2}} {\tau_\mathit{uv}}
         -\frac{e^{-M_n^{2}\tau_\mathit{ir}^2}}{\tau_\mathit{ir}} & 
\nonumber\\
+ \sqrt{\pi} M_{n} \Big({\rm erfc}(M_{n}\tau_\mathit{uv})-{\rm erfc}(M_{n}\tau_\mathit{ir})& \Big)
 \Bigg], 
 \label{PTRenew}  
\end{align} 
where $M_{n}=\sqrt{M^{2}+\omega^{2}_{n}}$ 
and
${\rm erfc}(x)$ is the complementary error function. 
From this we obtain the gap equation for the mass $M_\phi$ with twisted boundary conditions if we replace
$\omega_n$ by the shifted Matsubara frequencies $\omega_n^\phi$, given in Eq.~(\ref{omeganphi}).

Having solved the gap equations,
the chiral condensate is immediately obtained from Eq.~(\ref{pbpM}), which also holds at nonzero temperature.
In the same way we get the generalized condensate $\langle\bar{\psi}\psi\rangle_\phi$ as
\beq
        \langle\bar{\psi}\psi\rangle_\phi = -\frac{M_\phi-m}{G}
\label{pbpMphi}
\eeq
from the solution $M_\phi$ of the gap equation with twisted boundary conditions.
$\Sigma_1$ is then easily calculated from Eq.~(\ref{Sigman}) with $n=1$.

Finally, we would like to point out that the cancellation of the propagator poles in Eq.~(\ref{PTRc}) is independent
of the constituent quark mass. In particular, the quarks remain ``confined'' in the chirally restored phase where 
$M=0$.\footnote{
In order to ``cure'' this problem the authors of Ref.~\cite{Wang:2013wk}
introduced a temperature dependent IR cut-off $\tau_{ir}(T)$, which diverges in the chirally restored phase, 
so that chiral and deconfinement phase transition coincide.
}

\subsection{Nonlocal NJL model}
\label{sec:nNJL}

The nonlocal Nambu-Jona-Lasinio (nNJL) model is described by the Euclidean Lagrangian~\cite{Scoccola01}
\begin{equation}
\mathcal{L}_\mathrm{nNJL}  = \bar{\psi}(-i\slashed{\partial}+m)\psi-\frac{G}{2}j_a j_a,
\end{equation}
where $\psi$, $m$, and $G$ are again the quark field, its bare mass, and a coupling constant, respectively.
The nonlocality of the model is encoded within the nonlocal currents $j_a(x)$ defined as
\begin{equation}
j_a(x)=\int d^4y\,d^4z\;r(y-x)r(z-x)\,\bar{\psi}(y)\Gamma_a\psi(z),
\label{ja}
\end{equation}
with operators $\Gamma_a\in \{1,i\gamma^5\boldsymbol{\tau}\}$, again corresponding to 
the scalar-isoscalar and pseudoscalar-isovector channels.
The function $r(x)$ is a regulator,  which will be specified later.

In the local limit, $r(x) = \delta(x)$, the integrals in Eq.~(\ref{ja})
become trivial and we recover the standard NJL model,  Eq.~(\ref{LNJL}).
In general, however, the results get modified by the nonlocality.
Most important, the one-loop quark self-energy, which for local interactions
is constant in momentum space, 
now becomes a momentum dependent function. 
As a result the dressed propagator takes the form
\begin{equation}
S(q)=\frac{-\slashed{q}+\Sigma(q^2)}{q^2+\Sigma^2(q^2)},
\label{Sdressednl}
\end{equation}
where the function $\Sigma(q^2)$ replaces the constituent mass $M$ in Eq.~(\ref{Sdressed}).

For general nonlocal interactions the functional form of $\Sigma(q^2)$ must be found 
by self-consistently solving a Schwinger-Dyson equation. 
In the present model, however, a great simplification comes about from the fact that the
interaction is separable,
meaning that the four-point vertices in momentum space are essentially proportional to the 
product of the Fourier-transformed regulator functions. 
The form of the function $\Sigma(q^2)$ is then simply given by
\beq
        \Sigma(q^2)=m+\bar{\sigma}r^2(q^2) ,
\label{Sigmaform}
\eeq        
where  $r(q^2)$ is the regulator function in momentum space and $\bar{\sigma}$ is a constant,
satisfying the gap equation      
\beq
\bar{\sigma} = G\int\frac{d^4p}{(2\pi)^4}\, r^2(p^2)\tr(S(p)) .
\label{gapnl}
\eeq
Inserting Eqs.~(\ref{Sdressednl}) and (\ref{Sigmaform}) this becomes
\beq
\bar{\sigma}
= 8N_c G\int\frac{d^4p}{(2\pi)^4}\, r^2(p^2)\, 
\frac{m+\bar{\sigma}r^2(p^2)}{p^2+ (m+\bar{\sigma}r^2(p^2))^2} ,
\eeq
which, for a given $r(q^2)$, must be solved for  $\bar{\sigma}$.     

As before, the chiral condensate is given by Eq.~(\ref{qqgen}).
However, if we compare this equation with the gap equation~(\ref{gapnl}) we see that the latter
contains extra regulator functions in the integrand, so that in contrast to 
Eq.~(\ref{pbpM}) there is no simple relation between 
$\langle\bar{\psi}\psi\rangle$ and $\bar{\sigma}$
(except for $r(q^2) = \mathit{const.}$, corresponding to a local interaction).
Explicitly we have
\beq
\langle\bar{\psi}\psi\rangle
= -8N_c \int\frac{d^4p}{(2\pi)^4}\,  
\frac{m+\bar{\sigma}r^2(p^2)}{p^2+ (m+\bar{\sigma}r^2(p^2))^2} .
\label{pbpnl}
\eeq

Finite temperature effects as well as twisted boundary conditions
can again be incorporated through the substitutions Eqs.~(\ref{FTsub1}) -- (\ref{omeganphi})
in the expressions above.

If the  function $r(p^2)$ is chosen to drop off sufficiently fast at high values of $p^2$,
the integral in Eq.~(\ref{gapnl}) converges,
so that no further regularization of the model is needed.
Nevertheless, for the quark condensate, Eq.~(\ref{pbpnl}),
there remains a quadratic divergence due to the bare quark mass.
This problem also exists in QCD and requires a proper mass renormalization, see, e.g., Ref.~\cite{CSFM:09}. 
In order to avoid such complications, we will restrict our numerical studies to the chiral limit, $m=0$,
where the quark condensate is convergent.  

Unlike the local NJL model, the dressed propagator in the nNJL model has a nontrivial pole structure.
Depending on the model parameters and in particular the regulator function, the pole positions $q^2 = -\mathcal{M}^2$,
given by 
\beq
        \Big(q^2 + \Sigma^2(q^2) \Big)\Big|_{q^2 = -\mathcal{M}^2} = 0
\label{nNJLpoles}
\eeq
can be real or complex. In the latter case, assuming that $r(q^2)$ is real for real $q^2$, the poles appear in
complex conjugate pairs, which may be parametrized as
\begin{equation}
\mathcal{M}^2=M^2\pm iM\Gamma.
\label{int}
\end{equation}
Following a quasiparticle picture, $M$ can be interpreted as the constituent mass of the quark
and $\Gamma$ as its decay width~\cite{ Yo02,Loewe01,Yo01}.
In this manner, a complex pole corresponds to an unstable quasiparticle state,
which could be interpreted as a manifestation of confinement.\footnote{
This interpretation is probably too naive in various aspects. 
First, the underlying assumption is that the propagator can be Wick-rotated in the 
usual way, so that a pole at $q^2 = -\mathcal{M}^2$ in Euclidean space corresponds to a pole at $+\mathcal{M}^2$
in Minkowski space. However, this property is spoiled by the existence of the complex conjugate poles themselves.
Second, physical resonances have a cut along the real axis and complex poles are allowed only on the second Riemann sheet.
One may, however, turn this argument around:
Complex conjugate poles on the first Riemann sheet violate microcausality and positivity.
Therefore such poles do not correspond to physical particle states, which in turn could be interpreted as confinement.
}

An important aspect in this context is that the pole structure may change as a function of temperature
from ``confining'' to ``not confining''. 
Hence,  in contrast to the PT regularized NJL model, where the quarks are always confined,
the nNJL model allows us to study not only  the chiral phase transition, 
but also the deconfinement phase transition.

\section{Results}
\label{sec:results}

\subsection{Local NJL model with infrared cutoff}~\label{sec:contact}

We are now ready to present our results for the behavior of the chiral condensate and the dual quark condensate
as functions of the temperature. 
We begin with the PT regularized NJL model, introduced in Sec.~\ref{sec:PTRNJL}.
We consider the chiral limit, $m= 0$, and adopt the parameters of Ref.~\cite{Roberts:2011wy},\footnote{
The model of Ref.~\cite{Roberts:2011wy} uses a local four-point interaction with the quantum numbers of 
a heavy-gluon exchange, parametrized by an effective coupling parameter $m_G = 132$~MeV.
In the scalar-pseudoscalar channel this is Fierz equivalent to the Lagrangian~(\ref{LNJL}) with 
$G = \frac{2}{9} m_G^{-2}$.}
\begin{align}
       G &= 1.275\cdot 10^{-5}~\mathrm{MeV^{-2}}~, \\
       \tau_{ir} &= (240~\mathrm{MeV})^{-1}~, \\
       \tau_{uv} &= (905~\mathrm{MeV})^{-1}~,
\end{align}
which have been fitted to vacuum properties in the pion and rho-meson sector.
With these parameters the vacuum values of the constituent quark mass and
of the chiral condensate per flavor are given by $M=358$~MeV and
$\langle\bar{u}u\rangle^{1/3} = \langle\bar{d}d\rangle^{1/3} = -243$~MeV, respectively.

\begin{figure}[!htb]
\begin{center}
\includegraphics[scale=0.42]{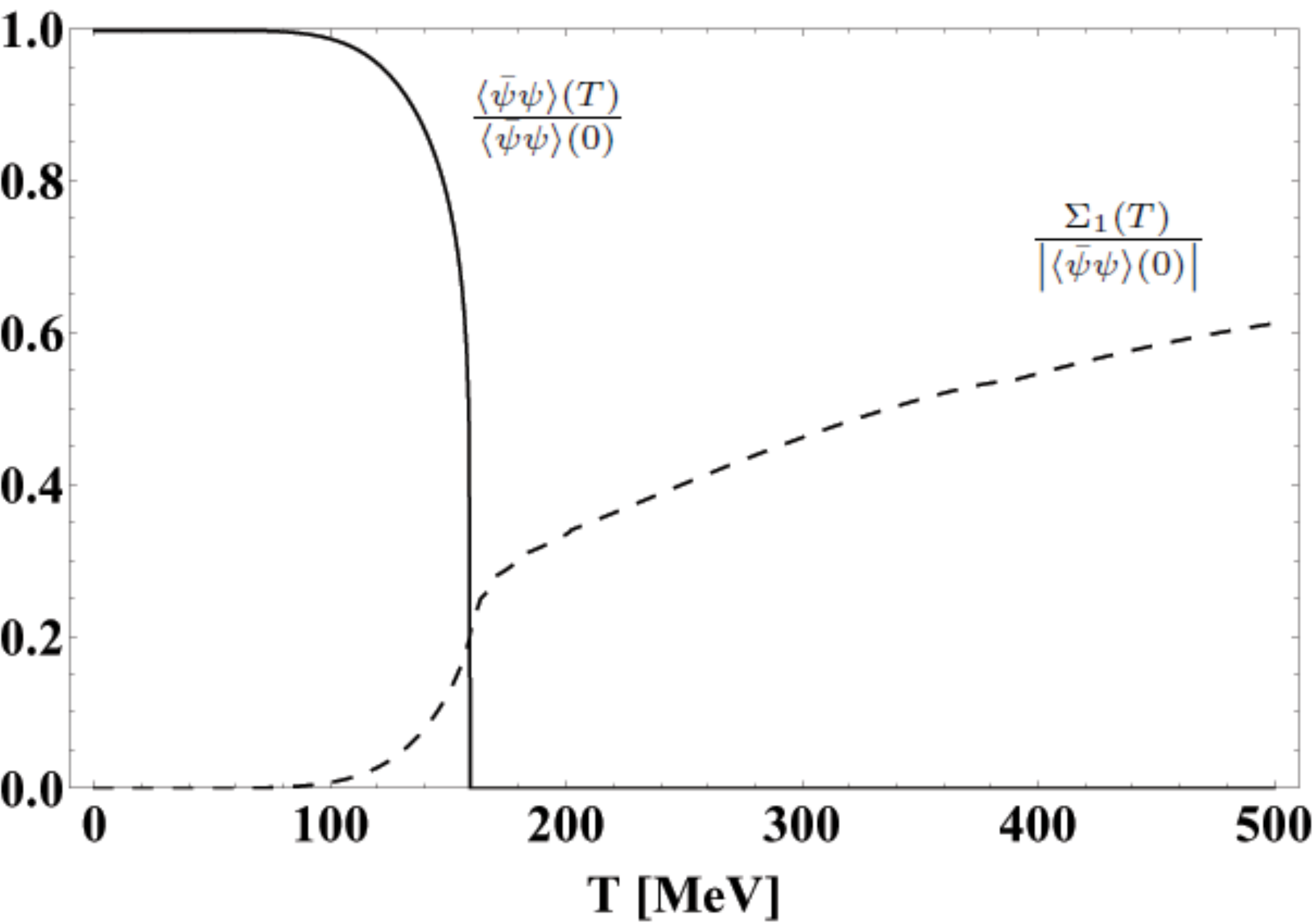}
\caption{
Absolute value of the quark condensate $ \langle\bar{\psi}\psi\rangle$ (solid line)
and  dual quark condensate $\Sigma_1$ (dashed line) as functions of the temperature,
calculated within the PT regularized NJL model in the chiral limit.
The condensates have been normalized by the absolute value of
$\langle\bar{\psi}\psi\rangle$ at $T=0$. 
}
\label{PTRDC:0}
\end{center}
\end{figure}

At nonzero temperature, we obtain the results displayed in Fig.~\ref{PTRDC:0}.
The solid and the dashed line indicate the absolute value of $\langle\bar{\psi}\psi\rangle$ and $\Sigma_1$, respectively,
both normalized by the absolute value of  $\langle\bar{\psi}\psi\rangle$ at $T=0$.
We find that $\langle\bar{\psi}\psi\rangle$ continuously goes to zero at a critical temperature $T_{c}=216.2$~MeV,
i.e., at this point a second-order chiral  phase transition takes place.
$\Sigma_1$, on the other hand, is very small at low temperatures and then smoothly rises,
with a maximum slope at $T=T_c$. 
In QCD this would indicate a (rather broad) crossover from the confined to the deconfined phase.
In the present model, however, the quarks remain confined, as discussed above. 
Hence, the rise of $\Sigma_1$ has nothing to do with a deconfinement transition in this case, 
but is simply triggered by the change of the chiral properties.

\subsection{Nonlocal NJL model}
\label{sec:nNJL}

For our investigations within the nNJL model we take a Gaussian regulator 
\begin{equation}
r^2(q)={\rm e}^{-q^2/\Lambda^2},
\end{equation}
which drops off exponentially at large Euclidean momenta, 
so that no further regularization is necessary.
Inserting this into Eqs.~(\ref{Sigmaform}) and (\ref{nNJLpoles})
one finds that the propagator has an infinite number of poles in the complex $q^2$ plane~\cite{Scoccola01}. 
However, depending on the model parameters, there may be poles at real $q^2$ as well. 
In the chiral limit, which will be considered in the following, 
the propagator has two real poles if the gap parameter $\bar\sigma$ is smaller than a critical 
value~\cite{Bowler:1994ir}
\beq
       \bar\sigma_\mathit{crit} = \frac{\Lambda}{\sqrt{2e}} ,
\label{sigmac}
\eeq
while there are no real poles if $\bar\sigma >  \bar\sigma_\mathit{crit}$.
As discussed before, the latter case can be interpreted as a manifestation of confinement. 

In our studies we therefore consider two qualitatively different sets of parameters,
which are listed in~Table \ref{t1}, together with the corresponding value of $\bar\sigma_\mathit{crit}$
and the vacuum solution $\bar\sigma_0$ of the gap parameter.
From these one can see that for parameter set A  the quarks in vacuum are confined (in the above sense),
while for set B they are not.
As temperature increases, $\bar\sigma$ decreases and finally vanishes at the chiral phase transition. 
Hence, for set A, there is also a deconfinement phase transition where  $\bar\sigma$ drops below 
$\bar\sigma_\mathit{crit}$.
For set B, on the other hand, the system is always in the deconfined phase.  
We are thus led to the question whether this qualitatively different behavior is also seen in the dual quark condensate.

\begin{table}[!htb]
\begin{tabular}{|c|c|c|c|c|}
\hline
Set & $\Lambda$~[MeV] &$G~[\mathrm{MeV}^{-2}]$ & $\bar{\sigma}_\mathit{crit}$~[MeV] & $\bar{\sigma}_0$~[MeV]\\
\hline
A&760&$3.6\cdot 10^{-5}$& 326 & 404 \\
B&914&$2.1\cdot 10^{-5}$&392 & 325 \\
\hline
\end{tabular}
\caption{Two sets of 
model parameters (regulator scale $\Lambda$ and coupling constant $G$), 
the corresponding critical gap parameter $\bar{\sigma}_\mathit{crit}$,
and the solution $\bar{\sigma}_0$ of the gap equation (\ref{gapnl}) at zero temperature.
The calculations are performed in the chiral limit, $m=0$.
}
\label{t1}
\end{table}

\begin{figure}[!htb]
\begin{center}
\includegraphics[scale=0.42]{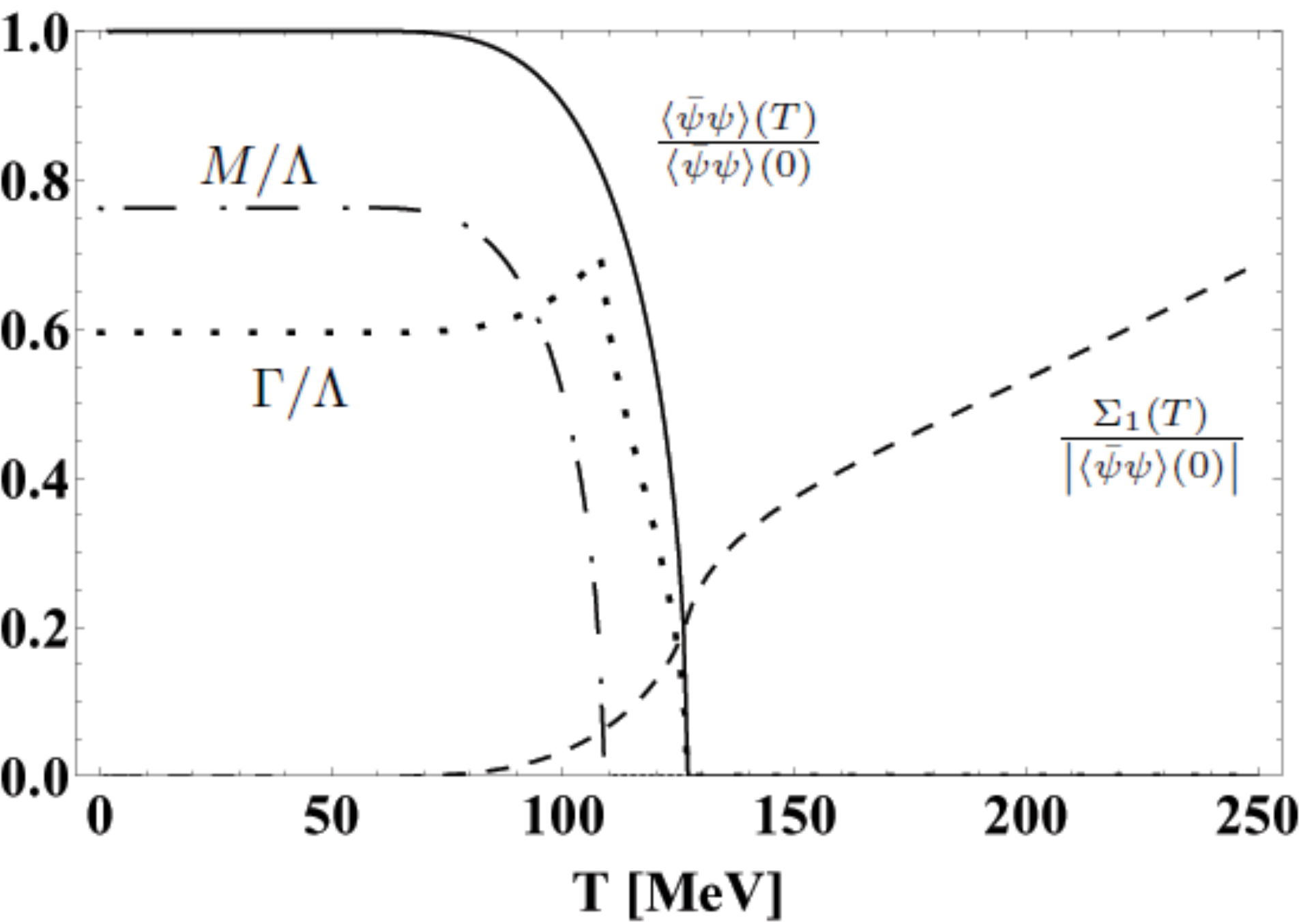}
\end{center}
\caption{Various quantities calculated with parameter set A as functions of the temperature:
Mass (dotted line) and decay width (dash-dotted line), according to Eq.~(\ref{int}) applied to the first propagator pole;
$\langle\bar{\psi}\psi\rangle$ (solid line), and $\Sigma_1$ (dashed line).  
}
\label{f3}
\end{figure}

Our results for parameter set A are displayed in Fig.~\ref{f3}.
Besides $\Sigma_1$ and the chiral condensate we also show the mass $M$ and the decay width $\Gamma$,  
according to the definition in Eq.~(\ref{int}). Here we focus on the propagator pole with the
lowest mass, since it contributes most significantly to the thermodynamics of the model.
We can see that the pole, which is complex at low temperatures becomes real at $T =110$~MeV,
which, according to the  interpretation discussed above, should be identified with the deconfinement temperature $T_d$.
The chiral phase transition, signaled by the vanishing of the mass and the chiral condensate,
takes place in the same regime but at a slightly higher temperature, $T_c  = 126.5$~MeV. 
This is of course expected because $\bar\sigma(T)$ first drops below 
$\bar\sigma_\mathit{crit}$  at $T=T_d$ before it vanishes completely at $T=T_c$. 

$\Sigma_1$ is again very small at low temperatures and then rises significantly. 
Like in the PT regularized local NJL model, the maximum slope is found at
the chiral transition temperature $T_c$.
However, since $\Sigma_1$ rises smoothly,  a relation to the deconfinement temperature $T_d$ cannot be totally excluded
from the figure. 

\begin{figure}[!htb]
\begin{center}
\includegraphics[scale=0.42]{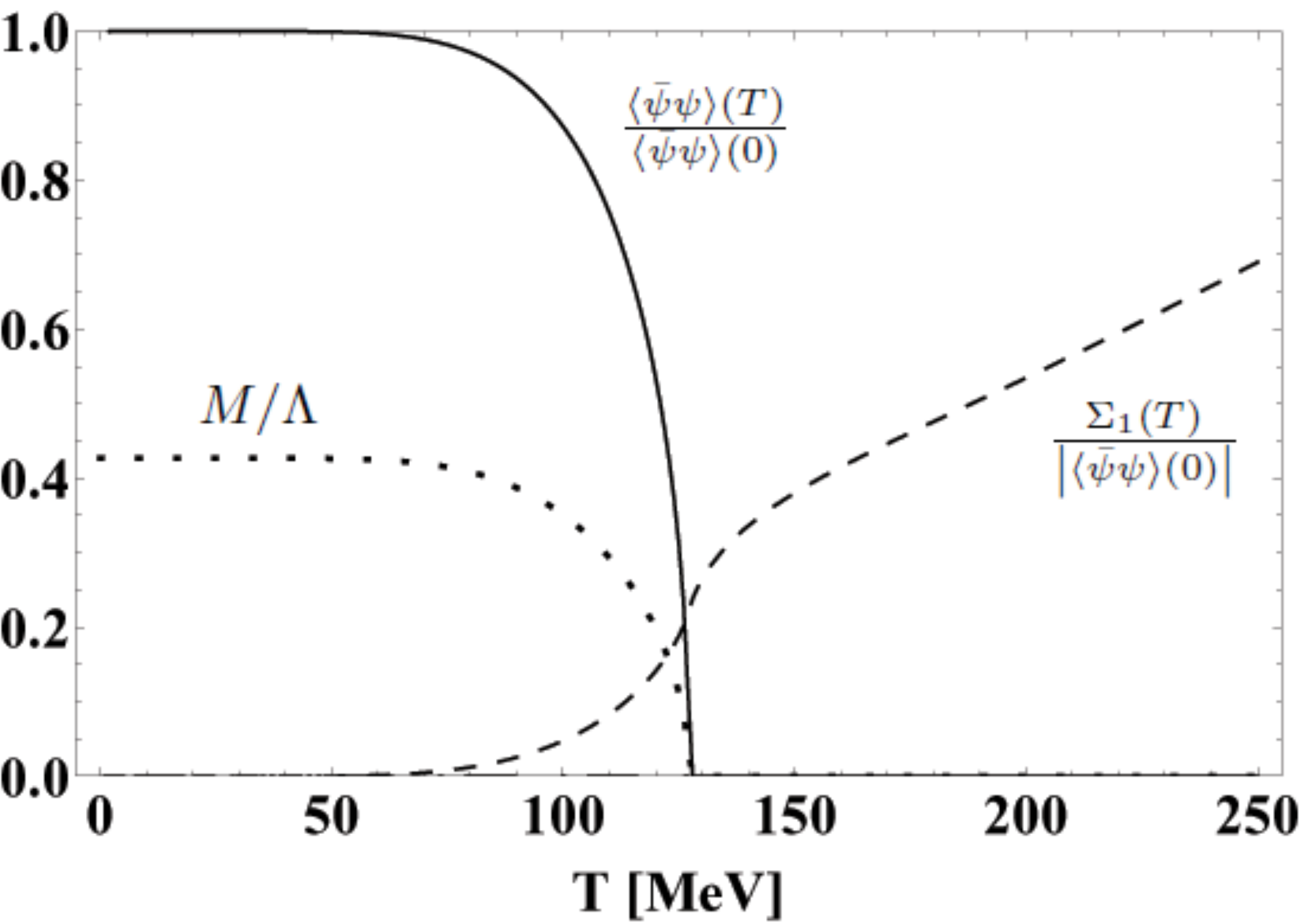}
\end{center}
\caption{The same as Fig.~\ref{f3}, but for parameter set B.
The decay width corresponding to the first propagator pole vanishes for all temperatures.
}
\label{f4}
\end{figure}

We therefore turn to the results for parameter set B, which are shown in Fig. \ref{f4}. 
Just as for parameter set A, the rise of $\Sigma_1$ occurs around the chiral phase transition, 
signaled by the vanishing of the chiral condensate. 
However, in this case there is no deconfinement phase transition occurring. 
Hence, the order parameter like behavior of the dual quark condensate can only be 
related to the chiral symmetry and should not be regarded as a sign of confinement in the model.

\section{Conclusions}~\label{sec:conclu}

In the present article we have studied the dual quark condensate $\Sigma_1$
in local and non-local variants of the NJL model. 
In QCD, $\Sigma_1$is an order parameter for confinement~\cite{EBF:08}, 
which is related to the breaking of the $\mathbb{Z}(3)$ center symmetry
and as such equivalent to the standard ``thin'' PL  in the limit of infinite quark masses.
In the NJL model, on the other hand, there is no center symmetry and no confinement,
but nevertheless $\Sigma_1$  behaves in a qualitatively similar way as in QCD~\cite{TKM:10}.
In Ref.~\cite{Benic01} this was shown within a Ginzburg-Landau type analysis
to be an effect of the chiral phase transition.

Like the original NJL model, the models studied in the present paper
do not have gauge fields and, hence, no center symmetry. 
However, there exists the possibility to have quark propagators without poles
on the real Euclidean $q^2$ axis, which is often interpreted as a realization of confinement. 
The aim of our analysis was therefore to investigate whether the pole structure of the quark propagator,
confining or non-confining in the above sense, leaves imprints on $\Sigma_1$. 

Specifically, we have considered three examples which all feature a chiral phase transition but 
have rather different confinement properties:
a PT regularized local NJL model with infrared cutoff, which is always confining,
a nonlocal NJL model with a deconfinement phase transition at finite temperature,
and a different parametrization of the same model where the quarks are always deconfined. 

We find that $\Sigma_1$ behaves almost identically in all three cases, 
namely it rises most steeply at the chiral phase transition temperature,
just like in the NJL model.
In particular, we do not see any effect related to a change of the confining properties of the propagator.
Although in one parametrization of the nNJL model there is a deconfinement transition 
and the rise of $\Sigma_1$ roughly falls in that region,
this must be seen as a coincidence because of the presence of the nearby chiral phase transition.
In fact, the two phase transitions do not take place at exactly at the same temperature, 
and the maximum slope of $\Sigma_1$ is found at the chiral rather than the deconfinement transition temperature.
Therefore we conclude that $\Sigma_1$ is not an appropriate order parameter for deconfinement in these models.

We would like to stress, however, that this  does {\it not} mean that the same conclusion can be drawn in QCD, 
where, unlike in the models we have studied, a connection between $\Sigma_1$ and center-symmetry breaking exists. 
Also, it is not clear whether confinement in QCD is really related to the pole structure of the quark propagator,
as assumed in our models. 
In any case, it seems that the connection between the rising behavior of $\Sigma_1$ and the chiral phase transition,
which in Ref.~\cite{Benic01} was shown for the NJL model, is a rather widespread feature.
Hence, if this is also true in QCD, it could explain the approximate coincidence of chiral and deconfinement 
crossovers, observed in lattice calculations.

\section{Acknowledgments}

We thank Adnan Bashir, Sanjin Benic and Christian Fischer for their comments on the manuscript. F.~M. acknowledges support from CONICYT (Chile) under grant No. 21110577. A.~A. and A.~R. acknowledge support from CONACyT (Mexico) and CIC-UMSNH. A.~A. also acknowledges the hospitality of PUC (Chile) and F. M. the hospitality of TU Darmstadt (Germany), where parts of this work were done.

\bibliography{dplbibnew}{}

\begin{thebibliography}{47}%
\makeatletter
\providecommand \@ifxundefined [1]{%
 \@ifx{#1\undefined}
}%
\providecommand \@ifnum [1]{%
 \ifnum #1\expandafter \@firstoftwo
 \else \expandafter \@secondoftwo
 \fi
}%
\providecommand \@ifx [1]{%
 \ifx #1\expandafter \@firstoftwo
 \else \expandafter \@secondoftwo
 \fi
}%
\providecommand \natexlab [1]{#1}%
\providecommand \enquote  [1]{``#1''}%
\providecommand \bibnamefont  [1]{#1}%
\providecommand \bibfnamefont [1]{#1}%
\providecommand \citenamefont [1]{#1}%
\providecommand \href@noop [0]{\@secondoftwo}%
\providecommand \href [0]{\begingroup \@sanitize@url \@href}%
\providecommand \@href[1]{\@@startlink{#1}\@@href}%
\providecommand \@@href[1]{\endgroup#1\@@endlink}%
\providecommand \@sanitize@url [0]{\catcode `\\12\catcode `\$12\catcode
  `\&12\catcode `\#12\catcode `\^12\catcode `\_12\catcode `\%12\relax}%
\providecommand \@@startlink[1]{}%
\providecommand \@@endlink[0]{}%
\providecommand \url  [0]{\begingroup\@sanitize@url \@url }%
\providecommand \@url [1]{\endgroup\@href {#1}{\urlprefix }}%
\providecommand \urlprefix  [0]{URL }%
\providecommand \Eprint [0]{\href }%
\providecommand \doibase [0]{http://dx.doi.org/}%
\providecommand \selectlanguage [0]{\@gobble}%
\providecommand \bibinfo  [0]{\@secondoftwo}%
\providecommand \bibfield  [0]{\@secondoftwo}%
\providecommand \translation [1]{[#1]}%
\providecommand \BibitemOpen [0]{}%
\providecommand \bibitemStop [0]{}%
\providecommand \bibitemNoStop [0]{.\EOS\space}%
\providecommand \EOS [0]{\spacefactor3000\relax}%
\providecommand \BibitemShut  [1]{\csname bibitem#1\endcsname}%
\let\auto@bib@innerbib\@empty
\bibitem [{\citenamefont {D{\"u}rr}\ \emph {et~al.}(2008)\citenamefont
  {D{\"u}rr}, \citenamefont {Fodor}, \citenamefont {Frison}, \citenamefont
  {Hoelbling}, \citenamefont {Hoffmann} \emph {et~al.}}]{Durr:2008zz}%
  \BibitemOpen
  \bibfield  {author} {\bibinfo {author} {\bibfnamefont {S.}~\bibnamefont
  {D{\"u}rr}}, \bibinfo {author} {\bibfnamefont {Z.}~\bibnamefont {Fodor}},
  \bibinfo {author} {\bibfnamefont {J.}~\bibnamefont {Frison}}, \bibinfo
  {author} {\bibfnamefont {C.}~\bibnamefont {Hoelbling}}, \bibinfo {author}
  {\bibfnamefont {R.}~\bibnamefont {Hoffmann}},  \emph {et~al.},\ }\href
  {\doibase 10.1126/science.1163233} {\bibfield  {journal} {\bibinfo  {journal}
  {Science}\ }\textbf {\bibinfo {volume} {322}},\ \bibinfo {pages} {1224}
  (\bibinfo {year} {2008})}\BibitemShut {NoStop}%
\bibitem [{\citenamefont {Borsanyi}\ \emph {et~al.}(2010)\citenamefont
  {Borsanyi} \emph {et~al.}}]{Borsanyi:2010bp}%
  \BibitemOpen
  \bibfield  {author} {\bibinfo {author} {\bibfnamefont {S.}~\bibnamefont
  {Borsanyi}} \emph {et~al.} (\bibinfo {collaboration} {Wuppertal-Budapest
  Collaboration}),\ }\href {\doibase 10.1007/JHEP09(2010)073} {\bibfield
  {journal} {\bibinfo  {journal} {JHEP}\ }\textbf {\bibinfo {volume} {1009}},\
  \bibinfo {pages} {073} (\bibinfo {year} {2010})}\BibitemShut {NoStop}%
\bibitem [{\citenamefont {Bazavov}\ \emph {et~al.}(2012)\citenamefont
  {Bazavov}, \citenamefont {Bhattacharya}, \citenamefont {Cheng}, \citenamefont
  {DeTar}, \citenamefont {Ding} \emph {et~al.}}]{Bazavov:2011nk}%
  \BibitemOpen
  \bibfield  {author} {\bibinfo {author} {\bibfnamefont {A.}~\bibnamefont
  {Bazavov}}, \bibinfo {author} {\bibfnamefont {T.}~\bibnamefont
  {Bhattacharya}}, \bibinfo {author} {\bibfnamefont {M.}~\bibnamefont {Cheng}},
  \bibinfo {author} {\bibfnamefont {C.}~\bibnamefont {DeTar}}, \bibinfo
  {author} {\bibfnamefont {H.}~\bibnamefont {Ding}},  \emph {et~al.},\ }\href
  {\doibase 10.1103/PhysRevD.85.054503} {\bibfield  {journal} {\bibinfo
  {journal} {Phys. Rev. D}\ }\textbf {\bibinfo {volume} {85}},\ \bibinfo
  {pages} {054503} (\bibinfo {year} {2012})}\BibitemShut {NoStop}%
\bibitem [{\citenamefont {Polyakov}(1978)}]{Polyakov:1978vu}%
  \BibitemOpen
  \bibfield  {author} {\bibinfo {author} {\bibfnamefont {A.~M.}\ \bibnamefont
  {Polyakov}},\ }\href {\doibase 10.1016/0370-2693(78)90737-2} {\bibfield
  {journal} {\bibinfo  {journal} {Phys. Lett.}\ }\textbf {\bibinfo {volume}
  {B72}},\ \bibinfo {pages} {477} (\bibinfo {year} {1978})}\BibitemShut
  {NoStop}%
\bibitem [{\citenamefont {Susskind}(1979)}]{Susskind:1979up}%
  \BibitemOpen
  \bibfield  {author} {\bibinfo {author} {\bibfnamefont {L.}~\bibnamefont
  {Susskind}},\ }\href {\doibase 10.1103/PhysRevD.20.2610} {\bibfield
  {journal} {\bibinfo  {journal} {Phys. Rev. D}\ }\textbf {\bibinfo {volume}
  {20}},\ \bibinfo {pages} {2610} (\bibinfo {year} {1979})}\BibitemShut
  {NoStop}%
\bibitem [{\citenamefont {McLerran}\ and\ \citenamefont
  {Svetitsky}(1981)}]{McLerran:1980pk}%
  \BibitemOpen
  \bibfield  {author} {\bibinfo {author} {\bibfnamefont {L.~D.}\ \bibnamefont
  {McLerran}}\ and\ \bibinfo {author} {\bibfnamefont {B.}~\bibnamefont
  {Svetitsky}},\ }\href {\doibase 10.1016/0370-2693(81)90986-2} {\bibfield
  {journal} {\bibinfo  {journal} {Phys. Lett.}\ }\textbf {\bibinfo {volume}
  {B98}},\ \bibinfo {pages} {195} (\bibinfo {year} {1981})}\BibitemShut
  {NoStop}%
\bibitem [{\citenamefont {Karsch}\ and\ \citenamefont
  {Wyld}(1985)}]{Karsch:1985cb}%
  \BibitemOpen
  \bibfield  {author} {\bibinfo {author} {\bibfnamefont {F.}~\bibnamefont
  {Karsch}}\ and\ \bibinfo {author} {\bibfnamefont {H.}~\bibnamefont {Wyld}},\
  }\href {\doibase 10.1103/PhysRevLett.55.2242} {\bibfield  {journal} {\bibinfo
   {journal} {Phys. Rev. Lett.}\ }\textbf {\bibinfo {volume} {55}},\ \bibinfo
  {pages} {2242} (\bibinfo {year} {1985})}\BibitemShut {NoStop}%
\bibitem [{\citenamefont {Bilgici}\ \emph {et~al.}(2008)\citenamefont
  {Bilgici}, \citenamefont {Bruckmann}, \citenamefont {Gattringer},\ and\
  \citenamefont {Hagen}}]{EBF:08}%
  \BibitemOpen
  \bibfield  {author} {\bibinfo {author} {\bibfnamefont {E.}~\bibnamefont
  {Bilgici}}, \bibinfo {author} {\bibfnamefont {F.}~\bibnamefont {Bruckmann}},
  \bibinfo {author} {\bibfnamefont {C.}~\bibnamefont {Gattringer}}, \ and\
  \bibinfo {author} {\bibfnamefont {C.}~\bibnamefont {Hagen}},\ }\href
  {\doibase 10.1103/PhysRevD.65.074021} {\bibfield  {journal} {\bibinfo
  {journal} {Phys. Rev. D}\ }\textbf {\bibinfo {volume} {77}},\ \bibinfo
  {pages} {094007} (\bibinfo {year} {2008})}\BibitemShut {NoStop}%
\bibitem [{\citenamefont {Braun}\ \emph {et~al.}(2011)\citenamefont {Braun},
  \citenamefont {Haas}, \citenamefont {Marhauser},\ and\ \citenamefont
  {Pawlowski}}]{Braun:2009gm}%
  \BibitemOpen
  \bibfield  {author} {\bibinfo {author} {\bibfnamefont {J.}~\bibnamefont
  {Braun}}, \bibinfo {author} {\bibfnamefont {L.~M.}\ \bibnamefont {Haas}},
  \bibinfo {author} {\bibfnamefont {F.}~\bibnamefont {Marhauser}}, \ and\
  \bibinfo {author} {\bibfnamefont {J.~M.}\ \bibnamefont {Pawlowski}},\ }\href
  {\doibase 10.1103/PhysRevLett.106.022002} {\bibfield  {journal} {\bibinfo
  {journal} {Phys. Rev. Lett.}\ }\textbf {\bibinfo {volume} {106}},\ \bibinfo
  {pages} {022002} (\bibinfo {year} {2011})}\BibitemShut {NoStop}%
\bibitem [{\citenamefont {Fischer}(2009)}]{CSF:09}%
  \BibitemOpen
  \bibfield  {author} {\bibinfo {author} {\bibfnamefont {C.~S.}\ \bibnamefont
  {Fischer}},\ }\href {\doibase 10.1103/PhysRevLett.103.052003} {\bibfield
  {journal} {\bibinfo  {journal} {Phys. Rev. Lett.}\ }\textbf {\bibinfo
  {volume} {103}},\ \bibinfo {pages} {052003} (\bibinfo {year}
  {2009})}\BibitemShut {NoStop}%
\bibitem [{\citenamefont {Fischer}\ and\ \citenamefont
  {Mueller}(2009)}]{CSFM:09}%
  \BibitemOpen
  \bibfield  {author} {\bibinfo {author} {\bibfnamefont {C.~S.}\ \bibnamefont
  {Fischer}}\ and\ \bibinfo {author} {\bibfnamefont {J.~A.}\ \bibnamefont
  {Mueller}},\ }\href {\doibase 10.1103/PhysRevD.65.074021} {\bibfield
  {journal} {\bibinfo  {journal} {Phys. Rev. D}\ }\textbf {\bibinfo {volume}
  {80}},\ \bibinfo {pages} {074029} (\bibinfo {year} {2009})}\BibitemShut
  {NoStop}%
\bibitem [{\citenamefont {Fischer}\ \emph {et~al.}(2010)\citenamefont
  {Fischer}, \citenamefont {Maas},\ and\ \citenamefont {M{\"u}ller}}]{CSFA:10}%
  \BibitemOpen
  \bibfield  {author} {\bibinfo {author} {\bibfnamefont {C.~S.}\ \bibnamefont
  {Fischer}}, \bibinfo {author} {\bibfnamefont {A.}~\bibnamefont {Maas}}, \
  and\ \bibinfo {author} {\bibfnamefont {J.~A.}\ \bibnamefont {M{\"u}ller}},\
  }\href {\doibase 10.1103/PhysRevD.65.074021} {\bibfield  {journal} {\bibinfo
  {journal} {Eur. Phys. J. C}\ }\textbf {\bibinfo {volume} {68}},\ \bibinfo
  {pages} {165} (\bibinfo {year} {2010})}\BibitemShut {NoStop}%
\bibitem [{\citenamefont {Fischer}\ \emph {et~al.}(2011)\citenamefont
  {Fischer}, \citenamefont {Luecker},\ and\ \citenamefont
  {Mueller}}]{Fischer:2011mz}%
  \BibitemOpen
  \bibfield  {author} {\bibinfo {author} {\bibfnamefont {C.~S.}\ \bibnamefont
  {Fischer}}, \bibinfo {author} {\bibfnamefont {J.}~\bibnamefont {Luecker}}, \
  and\ \bibinfo {author} {\bibfnamefont {J.~A.}\ \bibnamefont {Mueller}},\
  }\href {\doibase 10.1016/j.physletb.2011.07.039} {\bibfield  {journal}
  {\bibinfo  {journal} {Phys. Lett.}\ }\textbf {\bibinfo {volume} {B702}},\
  \bibinfo {pages} {438} (\bibinfo {year} {2011})}\BibitemShut {NoStop}%
\bibitem [{\citenamefont {Fischer}\ \emph {et~al.}(2014)\citenamefont
  {Fischer}, \citenamefont {Luecker},\ and\ \citenamefont
  {Welzbacher}}]{Fischer:2014ata}%
  \BibitemOpen
  \bibfield  {author} {\bibinfo {author} {\bibfnamefont {C.~S.}\ \bibnamefont
  {Fischer}}, \bibinfo {author} {\bibfnamefont {J.}~\bibnamefont {Luecker}}, \
  and\ \bibinfo {author} {\bibfnamefont {C.~A.}\ \bibnamefont {Welzbacher}},\
  }\href {\doibase 10.1103/PhysRevD.90.034022} {\bibfield  {journal} {\bibinfo
  {journal} {Phys. Rev. D}\ }\textbf {\bibinfo {volume} {90}},\ \bibinfo
  {pages} {034022} (\bibinfo {year} {2014})}\BibitemShut {NoStop}%
\bibitem [{\citenamefont {Kashiwa}\ \emph {et~al.}(2009)\citenamefont
  {Kashiwa}, \citenamefont {Kouno},\ and\ \citenamefont {Yahiro}}]{KKH:09}%
  \BibitemOpen
  \bibfield  {author} {\bibinfo {author} {\bibfnamefont {K.}~\bibnamefont
  {Kashiwa}}, \bibinfo {author} {\bibfnamefont {H.}~\bibnamefont {Kouno}}, \
  and\ \bibinfo {author} {\bibfnamefont {M.}~\bibnamefont {Yahiro}},\ }\href
  {\doibase 10.1103/PhysRevD.80.117901} {\bibfield  {journal} {\bibinfo
  {journal} {Phys. Rev. D}\ }\textbf {\bibinfo {volume} {80}},\ \bibinfo
  {pages} {117901} (\bibinfo {year} {2009})}\BibitemShut {NoStop}%
\bibitem [{\citenamefont {Mukherjee}\ \emph {et~al.}(2009)\citenamefont
  {Mukherjee}, \citenamefont {Chen},\ and\ \citenamefont {Huang}}]{TKM:10}%
  \BibitemOpen
  \bibfield  {author} {\bibinfo {author} {\bibfnamefont {T.~K.}\ \bibnamefont
  {Mukherjee}}, \bibinfo {author} {\bibfnamefont {H.}~\bibnamefont {Chen}}, \
  and\ \bibinfo {author} {\bibfnamefont {M.}~\bibnamefont {Huang}},\ }\href
  {\doibase 10.1103/PhysRevD.65.074021} {\bibfield  {journal} {\bibinfo
  {journal} {Phys. Rev. D}\ }\textbf {\bibinfo {volume} {82}},\ \bibinfo
  {pages} {034015} (\bibinfo {year} {2009})}\BibitemShut {NoStop}%
\bibitem [{\citenamefont {Nambu}\ and\ \citenamefont
  {Jona-Lasinio}(1961{\natexlab{a}})}]{Nambu:1961tp}%
  \BibitemOpen
  \bibfield  {author} {\bibinfo {author} {\bibfnamefont {Y.}~\bibnamefont
  {Nambu}}\ and\ \bibinfo {author} {\bibfnamefont {G.}~\bibnamefont
  {Jona-Lasinio}},\ }\href {\doibase 10.1103/PhysRev.122.345} {\bibfield
  {journal} {\bibinfo  {journal} {Phys. Rev.}\ }\textbf {\bibinfo {volume}
  {122}},\ \bibinfo {pages} {345} (\bibinfo {year}
  {1961}{\natexlab{a}})}\BibitemShut {NoStop}%
\bibitem [{\citenamefont {Nambu}\ and\ \citenamefont
  {Jona-Lasinio}(1961{\natexlab{b}})}]{Nambu:1961fr}%
  \BibitemOpen
  \bibfield  {author} {\bibinfo {author} {\bibfnamefont {Y.}~\bibnamefont
  {Nambu}}\ and\ \bibinfo {author} {\bibfnamefont {G.}~\bibnamefont
  {Jona-Lasinio}},\ }\href {\doibase 10.1103/PhysRev.124.246} {\bibfield
  {journal} {\bibinfo  {journal} {Phys. Rev.}\ }\textbf {\bibinfo {volume}
  {124}},\ \bibinfo {pages} {246} (\bibinfo {year}
  {1961}{\natexlab{b}})}\BibitemShut {NoStop}%
\bibitem [{\citenamefont {Feng}\ \emph {et~al.}(2013)\citenamefont {Feng},
  \citenamefont {Yin}, \citenamefont {Sun},\ and\ \citenamefont
  {Zong}}]{Feng:2013bsa}%
  \BibitemOpen
  \bibfield  {author} {\bibinfo {author} {\bibfnamefont {H.}~\bibnamefont
  {Feng}}, \bibinfo {author} {\bibfnamefont {P.}~\bibnamefont {Yin}}, \bibinfo
  {author} {\bibfnamefont {W.}~\bibnamefont {Sun}}, \ and\ \bibinfo {author}
  {\bibfnamefont {H.}~\bibnamefont {Zong}},\ }\href {\doibase
  10.1007/s11433-013-5084-7} {\bibfield  {journal} {\bibinfo  {journal} {Sci.
  China Phys. Mech. Astron.}\ }\textbf {\bibinfo {volume} {56}},\ \bibinfo
  {pages} {1116} (\bibinfo {year} {2013})}\BibitemShut {NoStop}%
\bibitem [{\citenamefont {Benic}(2013)}]{Benic01}%
  \BibitemOpen
  \bibfield  {author} {\bibinfo {author} {\bibfnamefont {S.}~\bibnamefont
  {Benic}},\ }\href {\doibase 10.1103/PhysRevD.88.077501} {\bibfield  {journal}
  {\bibinfo  {journal} {Phys. Rev. D}\ }\textbf {\bibinfo {volume} {88}},\
  \bibinfo {pages} {077501} (\bibinfo {year} {2013})}\BibitemShut {NoStop}%
\bibitem [{\citenamefont {Meisinger}\ and\ \citenamefont
  {Ogilvie}(1996)}]{Meisinger:1995ih}%
  \BibitemOpen
  \bibfield  {author} {\bibinfo {author} {\bibfnamefont {P.~N.}\ \bibnamefont
  {Meisinger}}\ and\ \bibinfo {author} {\bibfnamefont {M.~C.}\ \bibnamefont
  {Ogilvie}},\ }\href {\doibase 10.1016/0370-2693(96)00447-9} {\bibfield
  {journal} {\bibinfo  {journal} {Phys. Lett.}\ }\textbf {\bibinfo {volume}
  {B379}},\ \bibinfo {pages} {163} (\bibinfo {year} {1996})}\BibitemShut
  {NoStop}%
\bibitem [{\citenamefont {Fukushima}(2004)}]{Fukushima:2003fw}%
  \BibitemOpen
  \bibfield  {author} {\bibinfo {author} {\bibfnamefont {K.}~\bibnamefont
  {Fukushima}},\ }\href {\doibase 10.1016/j.physletb.2004.04.027} {\bibfield
  {journal} {\bibinfo  {journal} {Phys. Lett.}\ }\textbf {\bibinfo {volume}
  {B591}},\ \bibinfo {pages} {277} (\bibinfo {year} {2004})}\BibitemShut
  {NoStop}%
\bibitem [{\citenamefont {Hansen}\ \emph {et~al.}(2007)\citenamefont {Hansen},
  \citenamefont {Alberico}, \citenamefont {Beraudo}, \citenamefont {Molinari},
  \citenamefont {Nardi} \emph {et~al.}}]{Hansen:2006ee}%
  \BibitemOpen
  \bibfield  {author} {\bibinfo {author} {\bibfnamefont {H.}~\bibnamefont
  {Hansen}}, \bibinfo {author} {\bibfnamefont {W.}~\bibnamefont {Alberico}},
  \bibinfo {author} {\bibfnamefont {A.}~\bibnamefont {Beraudo}}, \bibinfo
  {author} {\bibfnamefont {A.}~\bibnamefont {Molinari}}, \bibinfo {author}
  {\bibfnamefont {M.}~\bibnamefont {Nardi}},  \emph {et~al.},\ }\href {\doibase
  10.1103/PhysRevD.75.065004} {\bibfield  {journal} {\bibinfo  {journal} {Phys.
  Rev. D}\ }\textbf {\bibinfo {volume} {75}},\ \bibinfo {pages} {065004}
  (\bibinfo {year} {2007})}\BibitemShut {NoStop}%
\bibitem [{\citenamefont {Krein}\ \emph {et~al.}(1992)\citenamefont {Krein},
  \citenamefont {Roberts},\ and\ \citenamefont {Williams}}]{Krein:1990sf}%
  \BibitemOpen
  \bibfield  {author} {\bibinfo {author} {\bibfnamefont {G.}~\bibnamefont
  {Krein}}, \bibinfo {author} {\bibfnamefont {C.~D.}\ \bibnamefont {Roberts}},
  \ and\ \bibinfo {author} {\bibfnamefont {A.~G.}\ \bibnamefont {Williams}},\
  }\href {\doibase 10.1142/S0217751X92002544} {\bibfield  {journal} {\bibinfo
  {journal} {Int. J. Mod. Phys.}\ }\textbf {\bibinfo {volume} {A7}},\ \bibinfo
  {pages} {5607} (\bibinfo {year} {1992})}\BibitemShut {NoStop}%
\bibitem [{\citenamefont {Burden}\ \emph {et~al.}(1992)\citenamefont {Burden},
  \citenamefont {Roberts},\ and\ \citenamefont {Williams}}]{Burden:1991gd}%
  \BibitemOpen
  \bibfield  {author} {\bibinfo {author} {\bibfnamefont {C.~J.}\ \bibnamefont
  {Burden}}, \bibinfo {author} {\bibfnamefont {C.~D.}\ \bibnamefont {Roberts}},
  \ and\ \bibinfo {author} {\bibfnamefont {A.~G.}\ \bibnamefont {Williams}},\
  }\href {\doibase 10.1016/0370-2693(92)91516-C} {\bibfield  {journal}
  {\bibinfo  {journal} {Phys. Lett.}\ }\textbf {\bibinfo {volume} {B285}},\
  \bibinfo {pages} {347} (\bibinfo {year} {1992})}\BibitemShut {NoStop}%
\bibitem [{\citenamefont {Stainsby}\ and\ \citenamefont
  {Cahill}(1992)}]{Stainsby:1992hy}%
  \BibitemOpen
  \bibfield  {author} {\bibinfo {author} {\bibfnamefont {S.}~\bibnamefont
  {Stainsby}}\ and\ \bibinfo {author} {\bibfnamefont {R.}~\bibnamefont
  {Cahill}},\ }\href {\doibase 10.1142/S0217751X92003410} {\bibfield  {journal}
  {\bibinfo  {journal} {Int.J.Mod.Phys.}\ }\textbf {\bibinfo {volume} {A7}},\
  \bibinfo {pages} {7541} (\bibinfo {year} {1992})}\BibitemShut {NoStop}%
\bibitem [{\citenamefont {Gribov}(1999)}]{Gribov:1999ui}%
  \BibitemOpen
  \bibfield  {author} {\bibinfo {author} {\bibfnamefont {V.}~\bibnamefont
  {Gribov}},\ }\href {\doibase 10.1007/s100529900052} {\bibfield  {journal}
  {\bibinfo  {journal} {Eur. Phys. J.}\ }\textbf {\bibinfo {volume} {C10}},\
  \bibinfo {pages} {91} (\bibinfo {year} {1999})}\BibitemShut {NoStop}%
\bibitem [{\citenamefont {Roberts}\ and\ \citenamefont
  {Schmidt}(2000)}]{Roberts:2000aa}%
  \BibitemOpen
  \bibfield  {author} {\bibinfo {author} {\bibfnamefont {C.~D.}\ \bibnamefont
  {Roberts}}\ and\ \bibinfo {author} {\bibfnamefont {S.~M.}\ \bibnamefont
  {Schmidt}},\ }\href {\doibase 10.1016/S0146-6410(00)90011-5} {\bibfield
  {journal} {\bibinfo  {journal} {Prog. Part. Nucl. Phys.}\ }\textbf {\bibinfo
  {volume} {45}},\ \bibinfo {pages} {S1} (\bibinfo {year} {2000})}\BibitemShut
  {NoStop}%
\bibitem [{\citenamefont {Alkofer}\ and\ \citenamefont {von
  Smekal}(2001)}]{Alkofer:2000wg}%
  \BibitemOpen
  \bibfield  {author} {\bibinfo {author} {\bibfnamefont {R.}~\bibnamefont
  {Alkofer}}\ and\ \bibinfo {author} {\bibfnamefont {L.}~\bibnamefont {von
  Smekal}},\ }\href {\doibase 10.1016/S0370-1573(01)00010-2} {\bibfield
  {journal} {\bibinfo  {journal} {Phys. Rept.}\ }\textbf {\bibinfo {volume}
  {353}},\ \bibinfo {pages} {281} (\bibinfo {year} {2001})}\BibitemShut
  {NoStop}%
\bibitem [{\citenamefont {Ebert}\ \emph {et~al.}(1996)\citenamefont {Ebert},
  \citenamefont {Feldmann},\ and\ \citenamefont {Reinhardt}}]{Ebert:1996vx}%
  \BibitemOpen
  \bibfield  {author} {\bibinfo {author} {\bibfnamefont {D.}~\bibnamefont
  {Ebert}}, \bibinfo {author} {\bibfnamefont {T.}~\bibnamefont {Feldmann}}, \
  and\ \bibinfo {author} {\bibfnamefont {H.}~\bibnamefont {Reinhardt}},\ }\href
  {\doibase 10.1016/0370-2693(96)01158-6} {\bibfield  {journal} {\bibinfo
  {journal} {Phys. Lett.}\ }\textbf {\bibinfo {volume} {B388}},\ \bibinfo
  {pages} {154} (\bibinfo {year} {1996})}\BibitemShut {NoStop}%
\bibitem [{\citenamefont {Hellstern}\ \emph {et~al.}(1997)\citenamefont
  {Hellstern}, \citenamefont {Alkofer},\ and\ \citenamefont
  {Reinhardt}}]{Hellstern:1997nv}%
  \BibitemOpen
  \bibfield  {author} {\bibinfo {author} {\bibfnamefont {G.}~\bibnamefont
  {Hellstern}}, \bibinfo {author} {\bibfnamefont {R.}~\bibnamefont {Alkofer}},
  \ and\ \bibinfo {author} {\bibfnamefont {H.}~\bibnamefont {Reinhardt}},\
  }\href {\doibase 10.1016/S0375-9474(97)00412-0} {\bibfield  {journal}
  {\bibinfo  {journal} {Nucl. Phys.}\ }\textbf {\bibinfo {volume} {A625}},\
  \bibinfo {pages} {697} (\bibinfo {year} {1997})}\BibitemShut {NoStop}%
\bibitem [{\citenamefont {Buballa}\ and\ \citenamefont
  {Krewald}(1992)}]{Buballa:1992sz}%
  \BibitemOpen
  \bibfield  {author} {\bibinfo {author} {\bibfnamefont {M.}~\bibnamefont
  {Buballa}}\ and\ \bibinfo {author} {\bibfnamefont {S.}~\bibnamefont
  {Krewald}},\ }\href {\doibase 10.1016/0370-2693(92)91632-J} {\bibfield
  {journal} {\bibinfo  {journal} {Phys. Lett.}\ }\textbf {\bibinfo {volume}
  {B294}},\ \bibinfo {pages} {19} (\bibinfo {year} {1992})}\BibitemShut
  {NoStop}%
\bibitem [{\citenamefont {Bowler}\ and\ \citenamefont
  {Birse}(1995)}]{Bowler:1994ir}%
  \BibitemOpen
  \bibfield  {author} {\bibinfo {author} {\bibfnamefont {R.}~\bibnamefont
  {Bowler}}\ and\ \bibinfo {author} {\bibfnamefont {M.}~\bibnamefont {Birse}},\
  }\href {\doibase 10.1016/0375-9474(94)00481-2} {\bibfield  {journal}
  {\bibinfo  {journal} {Nucl. Phys.}\ }\textbf {\bibinfo {volume} {A582}},\
  \bibinfo {pages} {655} (\bibinfo {year} {1995})}\BibitemShut {NoStop}%
\bibitem [{\citenamefont {Plant}\ and\ \citenamefont
  {Birse}(1998)}]{Plant:1997jr}%
  \BibitemOpen
  \bibfield  {author} {\bibinfo {author} {\bibfnamefont {R.~S.}\ \bibnamefont
  {Plant}}\ and\ \bibinfo {author} {\bibfnamefont {M.~C.}\ \bibnamefont
  {Birse}},\ }\href {\doibase 10.1016/S0375-9474(97)00635-0} {\bibfield
  {journal} {\bibinfo  {journal} {Nucl. Phys.}\ }\textbf {\bibinfo {volume}
  {A628}},\ \bibinfo {pages} {607} (\bibinfo {year} {1998})}\BibitemShut
  {NoStop}%
\bibitem [{\citenamefont {General}\ \emph {et~al.}(2001)\citenamefont
  {General}, \citenamefont {Gomez~Dumm},\ and\ \citenamefont
  {Scoccola}}]{General:2000zx}%
  \BibitemOpen
  \bibfield  {author} {\bibinfo {author} {\bibfnamefont {I.}~\bibnamefont
  {General}}, \bibinfo {author} {\bibfnamefont {D.}~\bibnamefont {Gomez~Dumm}},
  \ and\ \bibinfo {author} {\bibfnamefont {N.}~\bibnamefont {Scoccola}},\
  }\href {\doibase 10.1016/S0370-2693(01)00240-4} {\bibfield  {journal}
  {\bibinfo  {journal} {Phys. Lett.}\ }\textbf {\bibinfo {volume} {B506}},\
  \bibinfo {pages} {267} (\bibinfo {year} {2001})}\BibitemShut {NoStop}%
\bibitem [{\citenamefont {Loewe}\ \emph {et~al.}(2013)\citenamefont {Loewe},
  \citenamefont {Marquez},\ and\ \citenamefont {Villavicencio}}]{Yo01}%
  \BibitemOpen
  \bibfield  {author} {\bibinfo {author} {\bibfnamefont {M.}~\bibnamefont
  {Loewe}}, \bibinfo {author} {\bibfnamefont {F.}~\bibnamefont {Marquez}}, \
  and\ \bibinfo {author} {\bibfnamefont {C.}~\bibnamefont {Villavicencio}},\
  }\href {\doibase 10.1103/PhysRevD.88.056004} {\bibfield  {journal} {\bibinfo
  {journal} {Phys. Rev. D}\ }\textbf {\bibinfo {volume} {88}},\ \bibinfo
  {pages} {056004} (\bibinfo {year} {2013})}\BibitemShut {NoStop}%
\bibitem [{\citenamefont {Marquez}(2014)}]{Yo02}%
  \BibitemOpen
  \bibfield  {author} {\bibinfo {author} {\bibfnamefont {F.}~\bibnamefont
  {Marquez}},\ }\href@noop {} {\bibfield  {journal} {\bibinfo  {journal} {Phys.
  Rev. D}\ }\textbf {\bibinfo {volume} {89}},\ \bibinfo {pages} {076010}
  (\bibinfo {year} {2014})}\BibitemShut {NoStop}%
\bibitem [{\citenamefont {Alkofer}\ \emph {et~al.}(2004)\citenamefont
  {Alkofer}, \citenamefont {Detmold}, \citenamefont {Fischer},\ and\
  \citenamefont {Maris}}]{Alkofer:2003jj}%
  \BibitemOpen
  \bibfield  {author} {\bibinfo {author} {\bibfnamefont {R.}~\bibnamefont
  {Alkofer}}, \bibinfo {author} {\bibfnamefont {W.}~\bibnamefont {Detmold}},
  \bibinfo {author} {\bibfnamefont {C.}~\bibnamefont {Fischer}}, \ and\
  \bibinfo {author} {\bibfnamefont {P.}~\bibnamefont {Maris}},\ }\href
  {\doibase 10.1103/PhysRevD.70.014014} {\bibfield  {journal} {\bibinfo
  {journal} {Phys. Rev. D}\ }\textbf {\bibinfo {volume} {70}},\ \bibinfo
  {pages} {014014} (\bibinfo {year} {2004})}\BibitemShut {NoStop}%
\bibitem [{\citenamefont {Benic}\ \emph {et~al.}(2012)\citenamefont {Benic},
  \citenamefont {Blaschke},\ and\ \citenamefont {Buballa}}]{Benic:2012ec}%
  \BibitemOpen
  \bibfield  {author} {\bibinfo {author} {\bibfnamefont {S.}~\bibnamefont
  {Benic}}, \bibinfo {author} {\bibfnamefont {D.}~\bibnamefont {Blaschke}}, \
  and\ \bibinfo {author} {\bibfnamefont {M.}~\bibnamefont {Buballa}},\ }\href
  {\doibase 10.1103/PhysRevD.86.074002} {\bibfield  {journal} {\bibinfo
  {journal} {Phys. Rev. D}\ }\textbf {\bibinfo {volume} {86}},\ \bibinfo
  {pages} {074002} (\bibinfo {year} {2012})}\BibitemShut {NoStop}%
\bibitem [{\citenamefont {Vogl}\ and\ \citenamefont
  {Weise}(1991)}]{Vogl:1991qt}%
  \BibitemOpen
  \bibfield  {author} {\bibinfo {author} {\bibfnamefont {U.}~\bibnamefont
  {Vogl}}\ and\ \bibinfo {author} {\bibfnamefont {W.}~\bibnamefont {Weise}},\
  }\href {\doibase 10.1016/0146-6410(91)90005-9} {\bibfield  {journal}
  {\bibinfo  {journal} {Prog. Part. Nucl. Phys.}\ }\textbf {\bibinfo {volume}
  {27}},\ \bibinfo {pages} {195} (\bibinfo {year} {1991})}\BibitemShut
  {NoStop}%
\bibitem [{\citenamefont {Klevansky}(1992)}]{Klevansky01}%
  \BibitemOpen
  \bibfield  {author} {\bibinfo {author} {\bibfnamefont {S.}~\bibnamefont
  {Klevansky}},\ }\href {\doibase 10.1103/RevModPhys.64.649} {\bibfield
  {journal} {\bibinfo  {journal} {Rev. Mod. Phys.}\ }\textbf {\bibinfo {volume}
  {64}},\ \bibinfo {pages} {649} (\bibinfo {year} {1992})}\BibitemShut
  {NoStop}%
\bibitem [{\citenamefont {Hatsuda}\ and\ \citenamefont
  {Kunihiro}(1994)}]{Hatsuda:1994pi}%
  \BibitemOpen
  \bibfield  {author} {\bibinfo {author} {\bibfnamefont {T.}~\bibnamefont
  {Hatsuda}}\ and\ \bibinfo {author} {\bibfnamefont {T.}~\bibnamefont
  {Kunihiro}},\ }\href {\doibase 10.1016/0370-1573(94)90022-1} {\bibfield
  {journal} {\bibinfo  {journal} {Phys. Rept.}\ }\textbf {\bibinfo {volume}
  {247}},\ \bibinfo {pages} {221} (\bibinfo {year} {1994})}\BibitemShut
  {NoStop}%
\bibitem [{\citenamefont {Buballa}(2005)}]{Buballa01}%
  \BibitemOpen
  \bibfield  {author} {\bibinfo {author} {\bibfnamefont {M.}~\bibnamefont
  {Buballa}},\ }\href {\doibase 10.1016/j.physrep.2004.11.004} {\bibfield
  {journal} {\bibinfo  {journal} {Phys. Rept.}\ }\textbf {\bibinfo {volume}
  {407}},\ \bibinfo {pages} {205} (\bibinfo {year} {2005})}\BibitemShut
  {NoStop}%
\bibitem [{\citenamefont {Gomez~Dumm}\ and\ \citenamefont
  {Scoccola}(2002)}]{Scoccola01}%
  \BibitemOpen
  \bibfield  {author} {\bibinfo {author} {\bibfnamefont {D.}~\bibnamefont
  {Gomez~Dumm}}\ and\ \bibinfo {author} {\bibfnamefont {N.~N.}\ \bibnamefont
  {Scoccola}},\ }\href {\doibase 10.1103/PhysRevD.65.074021} {\bibfield
  {journal} {\bibinfo  {journal} {Phys. Rev. D}\ }\textbf {\bibinfo {volume}
  {65}},\ \bibinfo {pages} {074021} (\bibinfo {year} {2002})}\BibitemShut
  {NoStop}%
\bibitem [{\citenamefont {Wang}\ \emph {et~al.}(2013)\citenamefont {Wang},
  \citenamefont {Liu}, \citenamefont {Chang}, \citenamefont {Roberts},\ and\
  \citenamefont {Schmidt}}]{Wang:2013wk}%
  \BibitemOpen
  \bibfield  {author} {\bibinfo {author} {\bibfnamefont {K.-l.}\ \bibnamefont
  {Wang}}, \bibinfo {author} {\bibfnamefont {Y.-x.}\ \bibnamefont {Liu}},
  \bibinfo {author} {\bibfnamefont {L.}~\bibnamefont {Chang}}, \bibinfo
  {author} {\bibfnamefont {C.~D.}\ \bibnamefont {Roberts}}, \ and\ \bibinfo
  {author} {\bibfnamefont {S.~M.}\ \bibnamefont {Schmidt}},\ }\href {\doibase
  10.1103/PhysRevD.87.074038} {\bibfield  {journal} {\bibinfo  {journal} {Phys.
  Rev. D}\ }\textbf {\bibinfo {volume} {87}},\ \bibinfo {pages} {074038}
  (\bibinfo {year} {2013})}\BibitemShut {NoStop}%
\bibitem [{\citenamefont {Loewe}\ \emph {et~al.}(2011)\citenamefont {Loewe},
  \citenamefont {Morales},\ and\ \citenamefont {Villavicencio}}]{Loewe01}%
  \BibitemOpen
  \bibfield  {author} {\bibinfo {author} {\bibfnamefont {M.}~\bibnamefont
  {Loewe}}, \bibinfo {author} {\bibfnamefont {P.}~\bibnamefont {Morales}}, \
  and\ \bibinfo {author} {\bibfnamefont {C.}~\bibnamefont {Villavicencio}},\
  }\href {\doibase 10.1103/PhysRevD.83.096005} {\bibfield  {journal} {\bibinfo
  {journal} {Phys. Rev. D}\ }\textbf {\bibinfo {volume} {83}},\ \bibinfo
  {pages} {096005} (\bibinfo {year} {2011})}\BibitemShut {NoStop}%
\bibitem [{\citenamefont {Roberts}\ \emph {et~al.}(2011)\citenamefont
  {Roberts}, \citenamefont {Bashir}, \citenamefont {Gutierrez-Guerrero},
  \citenamefont {Roberts},\ and\ \citenamefont {Wilson}}]{Roberts:2011wy}%
  \BibitemOpen
  \bibfield  {author} {\bibinfo {author} {\bibfnamefont {H.}~\bibnamefont
  {Roberts}}, \bibinfo {author} {\bibfnamefont {A.}~\bibnamefont {Bashir}},
  \bibinfo {author} {\bibfnamefont {L.}~\bibnamefont {Gutierrez-Guerrero}},
  \bibinfo {author} {\bibfnamefont {C.}~\bibnamefont {Roberts}}, \ and\
  \bibinfo {author} {\bibfnamefont {D.}~\bibnamefont {Wilson}},\ }\href
  {\doibase 10.1103/PhysRevC.83.065206} {\bibfield  {journal} {\bibinfo
  {journal} {Phys. Rev. C}\ }\textbf {\bibinfo {volume} {83}},\ \bibinfo
  {pages} {065206} (\bibinfo {year} {2011})}\BibitemShut {NoStop}%
\end{thebibliography}%
\bibliographystyle{apsrev4-1}

\end{document}